\documentclass[english]{article}
\usepackage[T1]{fontenc}
\usepackage[latin1]{inputenc}
\usepackage{geometry}
\usepackage{graphicx}

\makeatletter

\newcommand{\lyxaddress}[1]{
\par {\center #1
\vspace{1.4em}
\noindent\par}
}

\usepackage{babel}
\makeatother
\begin{document}

\title{Towards determining the parameters of layer with scattering irregularities
that cause coherent echo, based on the Irkutsk Incoherent Scatter
radar data. }

\author{Grkovich K.V.,Berngardt O.I.}

\maketitle

\lyxaddress{Institute of Solar-Terrestrial physics SB RAS, Irkutsk, Russia}

\lyxaddress{E-mail: berng@iszf.irk.ru}

\begin{abstract}
In the paper we have presented a technique of determining the scattering
irregularities (that cause coherent echo) layer parameters using the
Irkutsk IS radar data. It is shown that our technique has necessary
accuracy (for height and thickness - about 2.5 km, for aspect sensitivity
- 5dB/degree). Processing of the experiments 25-26.12.1998 and 15-16.07.2000
has shown a good agreement of data calculated with the data obtained
by other investigators: an average layer height is 110-120km, average
layer thickness 5km, average aspect sensitivity - 15dB/degree. The
investigation of the experiments with high temporal resolution allowed
us to observe temporal variations of the irregularities layer parameters.
The average thickness and height of the layer does not contradict
the data obtained by other investigators. The investigation of the
experiments with high temporal resolution allowed to detect time variations
of the layer parameters. The temporal variations of the aspect sensitivity
are observed by us for the first time and requires additional investigations.
\end{abstract}

\section{{\normalsize Introduction.}}

One of the most important effects that affects to the ionospheric
and transionospheric radiowave propagation is a scattering on the
inhomogeneities of different scales. That is why, when studying the
ionosphere they pay valuable attention to the diagnostics of the ionospheric
inhomogeneities. At middle latitudes, the least observed but the most
affecting to the radiosignals propagation are the irregularities elongated
with the Earth magnetic field. Scattering by these irregularities
has high aspect sensitivity, that causes a strong dependence of scattered
signal amplitude on transmitted and received waves orientation. One
of the physical mechanisms of creation of such inhomogeneities, elongated
with the Earth magnetic field, in the ionospheric E-layer is the growth
of two-stream and gradient-drift irregularities. The necessary conditions
for the growth of these irregularities are powerful electric field,
high electron-to-ion velocity and high electron density gradients
existence \cite{Farley63,Buneman63}. Such inhomogeneities are observed
at high and equatorial latitudes and could affect to the received
signal from 8MHz to 1GHz \cite{Haldoupis89}. The conditions for arising
these inhomogeneities at mid-latitudes most frequently are satisfied
during strong geomagnetic disturbances. At polar and equatorial latitudes
these requirements are satisfied in less disturbed conditions \cite{Haldoupis89,st-Mourice89}.
Radiosignals scattered at these inhomogeneities are known as radioaurora,
or coherent echo (CE).The CE signals have been observed at the Irkutsk
IS radar since 1998. During the last time at the Irkutsk IS radar
a number of strong geomagnetic disturbances accompanied with CE observations
have been observed \cite{Zolotuhina07}.

The experiment geometry is shown at fig.1. Irkutsk IS radar is a monostatic
tool and transmits a periodical sequence of radiopulses. It receives
scattered signal after each transmitted pulse and averages scattered
power from pulse to pulse. Wide antenna patter allows us to analyze
the signal scattered from a big range of radar ranges almost at the
same time with a good time and spatial resolution (1-2 minutes and
15 km. correspondingly). More detailed the experiment geometry and
the radar characteristics was described at \cite{Potekhin99}.

~

\includegraphics[scale=0.6]{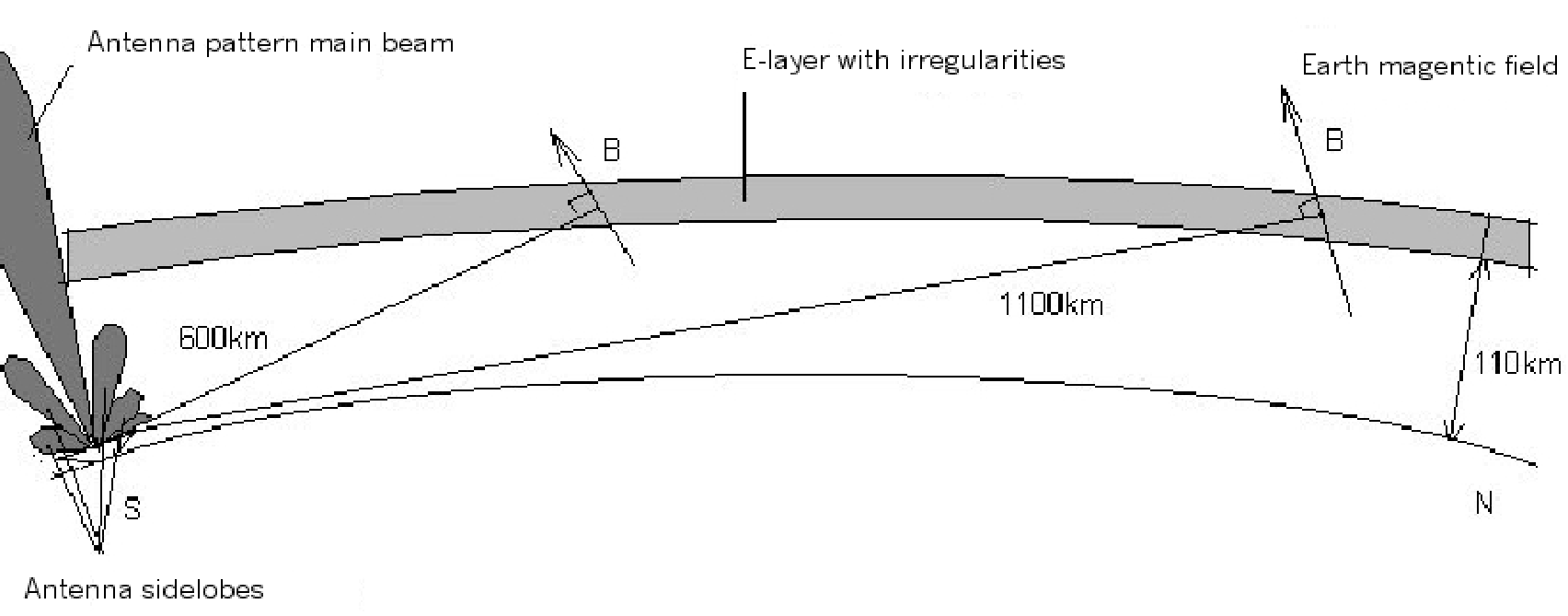}

{\small Fig 1. The geometry of the experiment}{\small \par}

~

Due to the Irkutsk IS radar location and construction peculiarities
\cite{Zherbtsov2002} the scattered CE signal usually has two peaks
in the power profile - at ranges 550 and 1100 km correspondingly,
at these ranges the line-of-sight is most close to the perpendicular
to the magnetic field. As preliminary analysis has shown the peaks
shape and location depend on the characteristics of scattering irregularities.
The aim of this paper is to determine scattering irregularities characteristics
within the therm of a standard model. The task is not a new one. Such
experiments have been made at some radars with narrow (or steerable)
antenna pattern \cite{st-Mourice89,Foster92}. This allow to determine
characteristics of scattering irregularities in terms of layer of
irregularities - height of the layer is approximately 110km, layer
thickness is about 10km, sometimes a number of thin layers has been
observed at the same time \cite{st-Mourice89}. The aspect sensitivity
has been investigated at these frequencies by other investigators
and gives average estimates of about 15 dB/degree \cite{Foster92}.
Due to the Irkutsk IS radar antenna pattern peculiarities the techniques
used at other radars do not work - antenna pattern at lower side lobes
(used to receive CE signals) is close to an isotropic one. That is
why the aim of the paper is to obtain a stable algorithm for determining
the characteristics of the layer with irregularities - the height,
the thickness and the aspect sensitivity of irregularities from the
power profile dependence on range.

\section{{\normalsize The model of the scattered signal.}}

For solving the declared aim the numerical modelling has been made
to obtain a dependence of CE power profile on radar range and scattering
layer characteristics. The observed power profile is produced by scattering
on irregularities generated during strong geomagnetic disturbances
in ionospheric E-layer and elongated with the Earth magnetic field. 

The model (\ref{eq:1}) used to describe average power of the scattered
signal on radar range (power profile $P(R)$) includes aspect sensitivity
of scattered signal in exponential form \cite{Foster92} (with accuracy
of an arbitrary multiplier):

\begin{equation}
P(R)=f(E)\int|u(2\frac{R-r}{c})|^{2}|g(\frac{\overrightarrow{r}}{r})|^{2}exp(-\frac{(h-h_{0})^{2}}{(\bigtriangleup h)^{2}})exp(-(\frac{arccos(\frac{\overrightarrow{r}\overrightarrow{B}}{rB})}{\bigtriangleup\varphi})^{2})\frac{d\overrightarrow{r}}{r^{2}}\label{eq:1}\end{equation}

$f(E)$- dependence on the electric field;

$u(2\frac{R-r}{c})$-sounding signal pulse shape;

$g(\frac{\overrightarrow{r}}{r})$-antenna pattern;

$h$- height at the scattering point;

$h_{0}$- layer height, $\bigtriangleup h$-layer thickness, $\bigtriangleup\varphi$-
aspect sensitivity;

$\overrightarrow{B}$- vector of magnetic field;

$\overrightarrow{r}$- vector from the observation point to the scattering
point.

The calculation has been made is based on the international reference
model for magnetic field (IGRF) and the Irkutsk IS radar antenna pattern
model (currently used by us) $g(\frac{\overrightarrow{r}}{r})$. As
a model of scattering irregularities a Gaussian layer with irregularities
has been chosen, characterized by its height $h_{0}$ and thickness
$\bigtriangleup h$. Spectral density of the irregularities, which
defines the aspect sensitivity of the scattered signal power from
the magnetic field was chosen in exponential form. This dependence
is defined by the aspect sensitivity $\bigtriangleup\varphi$ and
by the angle between line-of-sight and magnetic field $\overrightarrow{B}$.
The electric field $E$ effects in the scattered power is included
as some multiplier $f(E)$. The shape of the transmitted pulse $u(t)$
was also taken into consideration. The coordinate center $\overrightarrow{r}=0$
corresponds to the antenna center.

At fig.1 power profiles, which were obtained by the described model
for different parameters $h_{0},\bigtriangleup h,\bigtriangleup\varphi$,
are shown. The red-color line corresponds to the average parameters
of irregularities. From this figure one can see that profiles for
the different parameters can change drastically. 

~

\includegraphics[scale=0.6, angle=270]{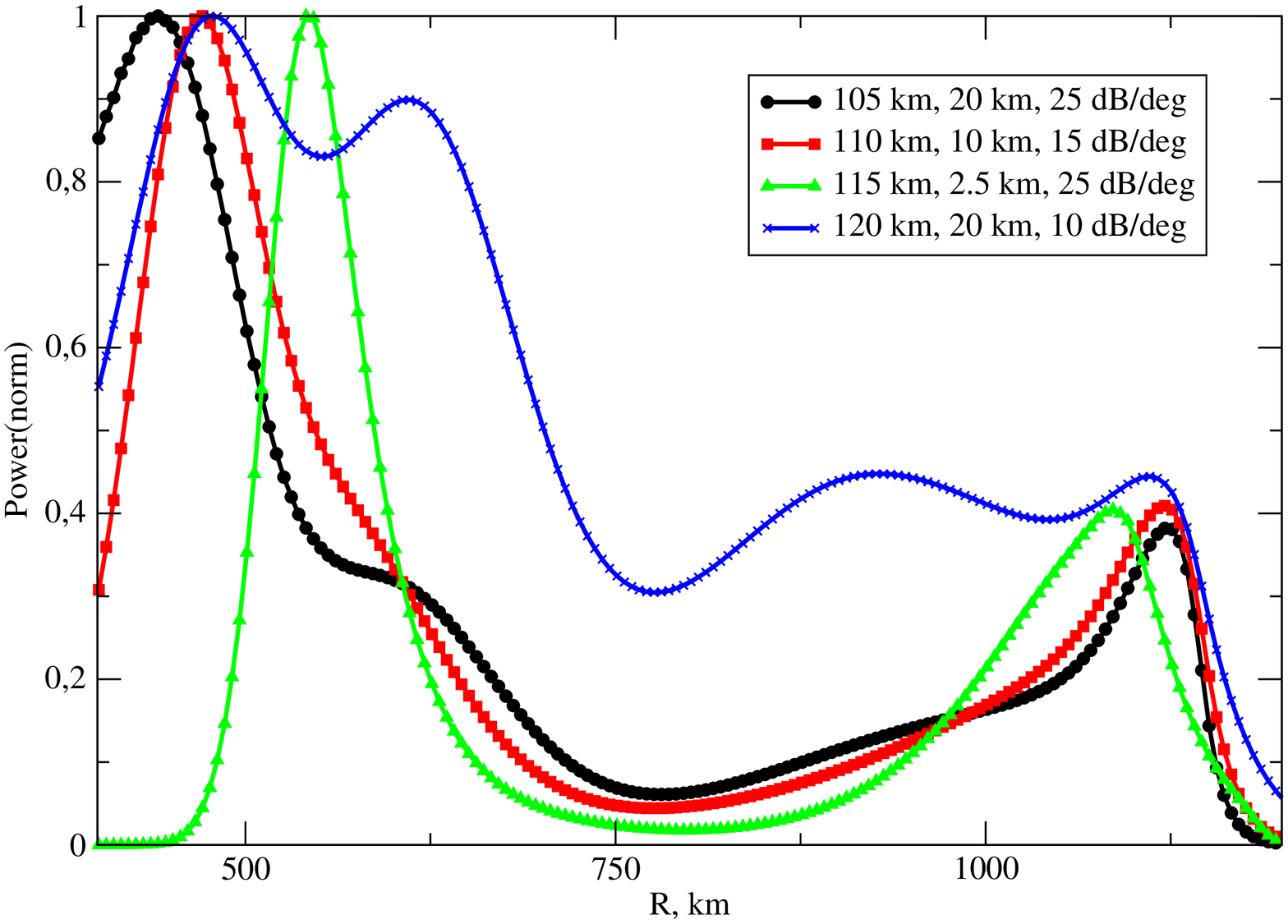}

{\small Fig.2 Model power profiles for the different model values}{\small \par}

~

The search of unknown parameters $h_{0},\bigtriangleup h,\bigtriangleup\varphi$
over the experimentally measured power profiles was made by the search
of the correlation coefficient over the number of preliminary calculated
model profiles (\ref{eq:2}). The indexes $mod$ and $exp$ for power
profiles corresponds to the power profiles calculated from the model
and for experimental power profiles correspondingly. The accumulation
$\begin{array}{c}
\Sigma\\
i\end{array}$ should be made over the radar range discreets $P_{i}=P(R_{i},h_{0},\bigtriangleup h,\bigtriangleup\varphi)$. 

\begin{equation}
K(h_{0},\bigtriangleup h,\bigtriangleup\varphi)=\frac{\begin{array}{c}
\Sigma\\
i\end{array}(P_{i}^{exp}*(P_{i}^{mod}+C))}{\sqrt{\begin{array}{c}
\Sigma\\
i\end{array}(P_{i}^{exp})*\begin{array}{c}
\Sigma\\
i\end{array}(P_{i}^{mod}+C)^{2}}}\label{eq:2}\end{equation}

The search of the constant $C$ which defines the noise level (ionospheric
and electronical noise) was made according to the functional (\ref{eq:2})
maximum condition in form (\ref{eq:3}). It could be rewritten in
form of the linear equation for $C$ (\ref{eq:4}):

\begin{equation}
\frac{dK}{dC}=0\label{eq:3}\end{equation}

\begin{equation}
C=\frac{\begin{array}{c}
\Sigma\\
i\end{array}P_{i}^{exp}*\begin{array}{c}
\Sigma\\
i\end{array}(P_{i}^{mod})^{2}-\begin{array}{c}
\Sigma\\
i\end{array}(P_{i}^{exp}*P_{i}^{mod})*\begin{array}{c}
\Sigma\\
i\end{array}P_{i}^{mod}}{n*\begin{array}{c}
\Sigma\\
i\end{array}(P_{i}^{exp}*P_{i}^{mod})-\begin{array}{c}
\Sigma\\
i\end{array}P_{i}^{mod}*\begin{array}{c}
\Sigma\\
i\end{array}P_{i}^{exp}}\label{eq:4}\end{equation}

The range of layer height values was chosen from 100 to 125 km, layer
thickness - from 2.5 to 20 km, aspect sensitivity - from10 t0 25dB/grad.
With the step of 2.5 km for height and thickness, and 2.5dB/grad for
aspect sensitivity the model power profiles have been calculated and
summarized as a matrix. To check the stability of the solution, including
the presence of the noise, a number of tests has been made to test
its stability at all model parameters combinations. The testing has
been made by adding noises which correspond both to the ionospheric
noise and statistical noise which arise due to limited number of samples
used for averaging. During the testing 3 variants of the noise have
been used. Averaging has been made over 500 single soundings:

\begin{enumerate}
\item The noise with constant average amplitude that corresponds to the
ionospheric noise. In this case the noise amplitude was 2\% from the
power profile maximum.
\item The noise with amplitude that is proportional to the power signal
at given radar range, it corresponds to the statical noises. The amplitude
of the noise was 4\% from the current power profile value (at given
height).
\item Low-frequency noise, corresponding to the receiving technique causes.
Noise amplitude was 2\% from the power profile maximum, frequency
was within actual filters bandwidth used for experiments.
\end{enumerate}
The error distributions for testing results are shown at fig.3. Green
color corresponds to the low-frequency noise, blue color corresponds
to the ionospheric-like noise, magenta color corresponds to the statistical
noise. At the left top of the figure - the results for the layer height,
at the right top - for the layer thickness, at the bottom - for the
aspect sensitivity. From this figure it is clear that aspect sensitivity
is less stable parameter to calculate, the most stable parameters
are layer height and thickness. When calculating aspect sensitivity,
the algorithm provides less stable results, but inspite of this it
gives a good precision, and errors with amplitudes more than model
discreet 2.5dB/degree do not exceed 2\% of all the tested samples.
The algorithm provides less stability for low-frequency noises.

\includegraphics[scale=0.6, angle=270]{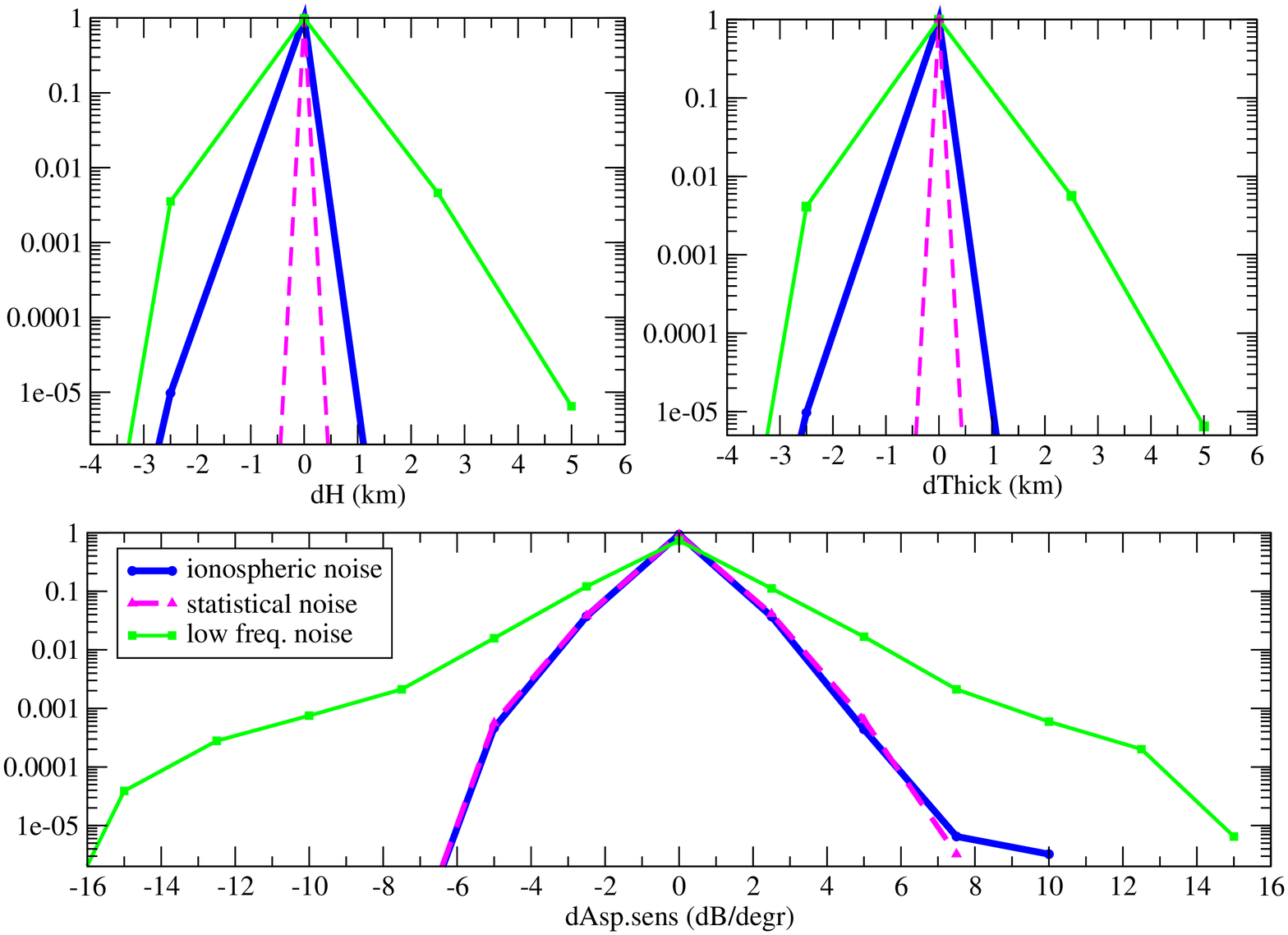}

{\small Fig.3 Errors distribution of inverting of noised model data.}{\small \par}

~

In addition to testing the model inversion accuracy in presence of
the noise, a testing of the experimental data was made in presence
of the same noises. The obtained results differ from the results obtained
when inverting the model profiles in presence of the noise. The most
stable parameters are still layer height and thickness. The less stable
is aspect sensitivity, that was calculated with unsuccessful accuracy.
As opposed to the testing of the model data, the testing of the experimental
data provides non-symmetrical error distributions, and in number of
cases does not provide the necessary accuracy. This effect could be
explained by the fact that a number of cases we could not define as
'coherent echo'. Inspite of this the number of parameters erroneously
defined (that are out of model accuracy) is pretty small, and this
allows us to suggest the model as adequate to the experiment data.

\includegraphics[scale=0.6, angle=270]{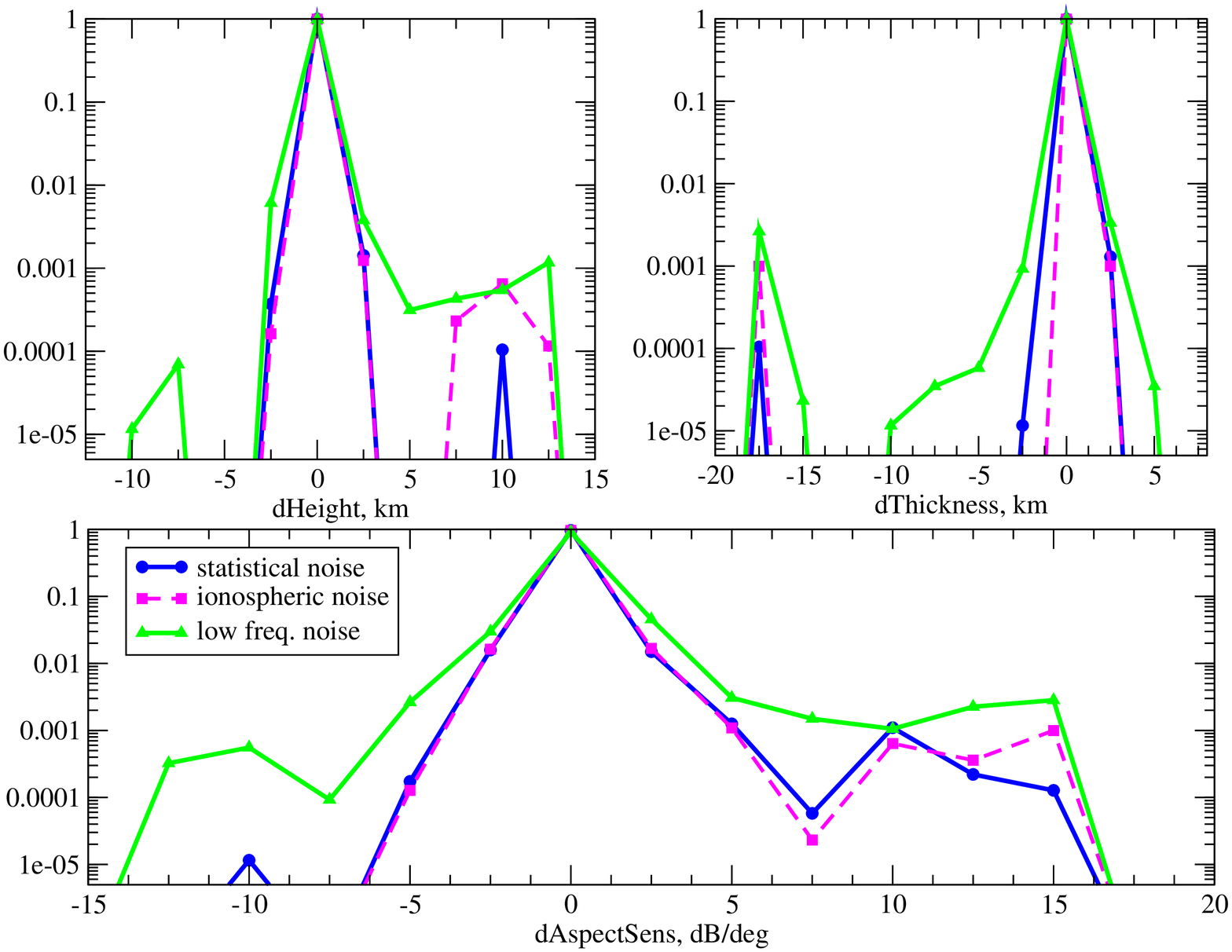}

{\small Fig.4a Errors distribution of inverting noised experimental
data ( 25-26.09.1998 experiment)}{\small \par}

~

\includegraphics[scale=0.6, angle=270]{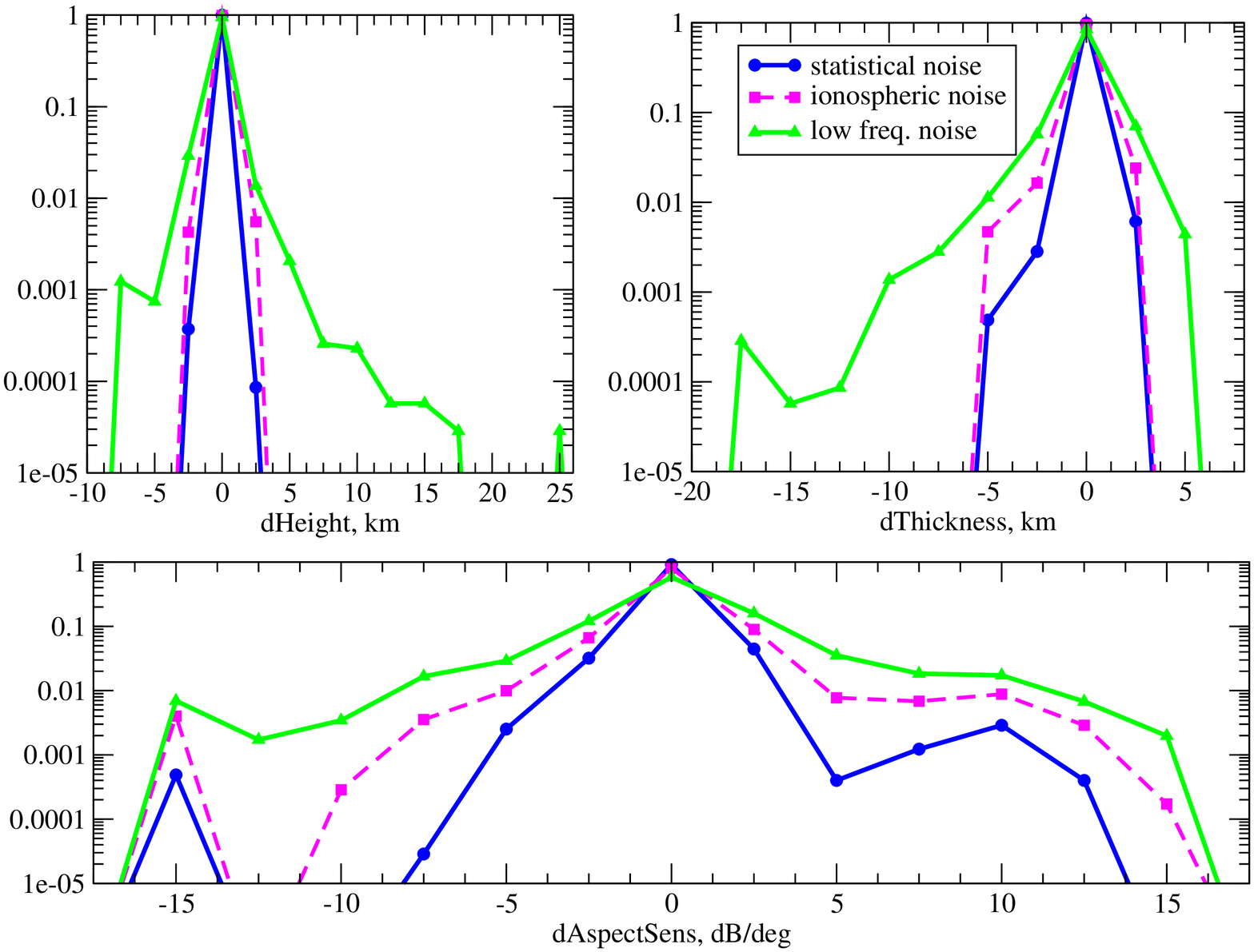}

{\small Fig 4b. Errors distribution of inverting noised experimental
data ( 16.07.2000 experiment)}{\small \par}

\section{{\normalsize Discussion.}}

During the work the data obtained at Irkutsk IS radar was processed
for strong geomagnetic storms on 25-26.09.1998 and 15-16.07.2000.
The data processing has shown a successful agreement between the model
and the experimental data (more than a half of the experimental profiles
agrees with the experimental data, with the correlation coefficient
0.98, and in 90\% of coherent echo observation cases the correlation
coefficient is no less than 0.94). At fig.5a-b the dependence of the
parameters (inverted with the described technique) on time is shown.
From the fig.5a-b one can see that time dependence of irregularities
layer parameters (especially its height and thickness) usually has
smooth time variations. So we can explain power profile time dependence
by variations of these parameters. This allows us to suppose the correctness
of the model for description of the experimental data.

\includegraphics[scale=0.6, angle=270]{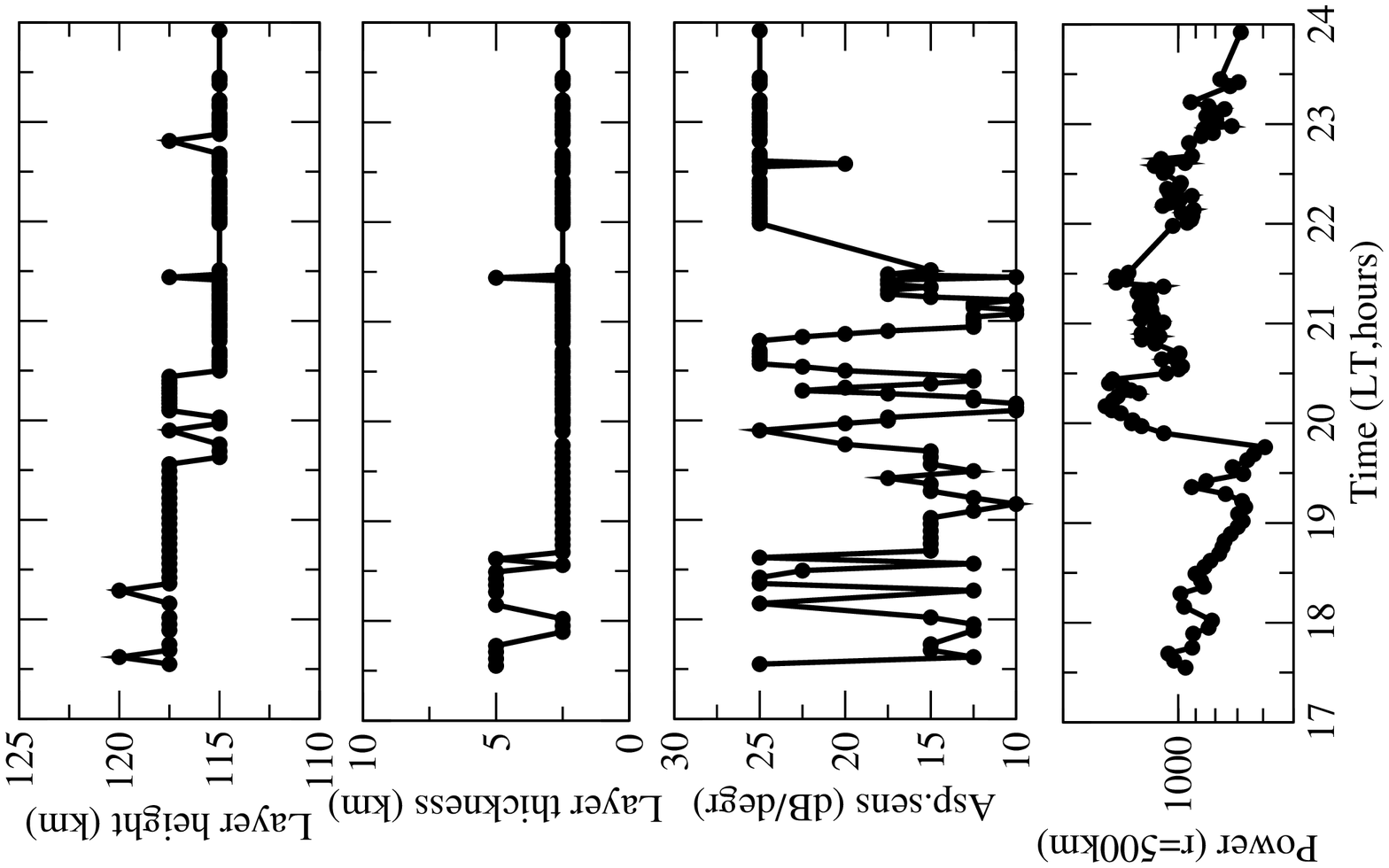}

{\small Fig 5a Experimental data processing results for 25-26.09.1998
experiment.}{\small \par}

~

\includegraphics[scale=0.6, angle=270]{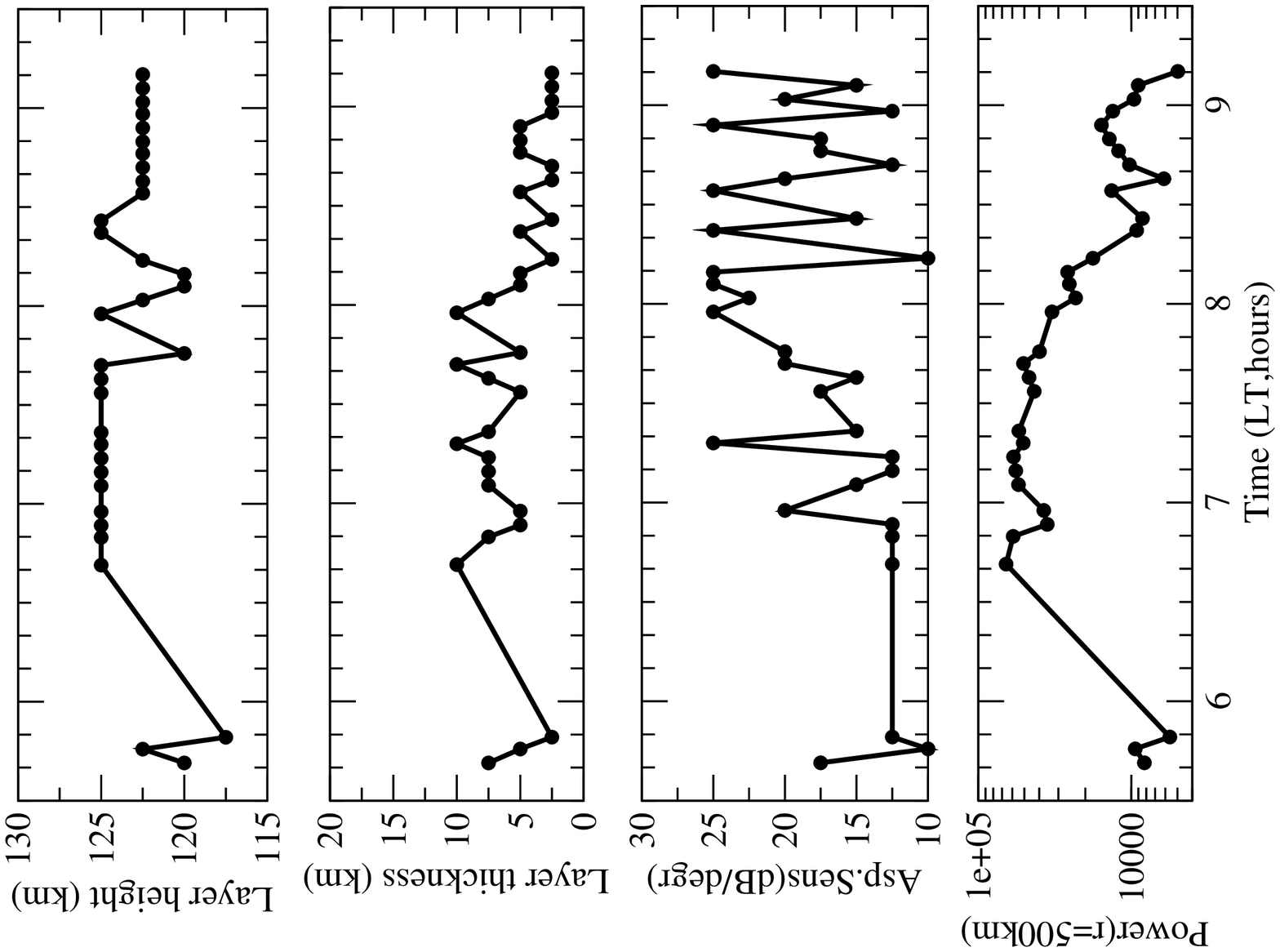}

{\small Fig 5b Experimental data processing results for 15-16.07.2000
experiment.}{\small \par}

~

The fast variations observed especially in aspect sensitivity time
variations could be caused by the following:

1. Non-optimal time resolution of the experimental data (for now \textasciitilde{}
4 minutes)

2. Latitudinal changes of the model parameters, not included into
consideration. It is the electron-ion drift velocity, that could cause
the most signifficant changes of any parameter.

3. Errors in the Irkutsk IS radar antenna pattern model (current model
has not been calibrated at lower sidelobes, affecting to the power
profile shape) 

4. Not including into consideration an azimuthal aspect sensitivity
of the irregularities (the model of scattered signal is isotropic
in the plane perpendicular to the magnetic field, due to the fact
that we do not know electric field direction exactly).

\section{{\normalsize Conclusion}}

In the paper we have presented a technique of determining the scattering
irregularities (that cause coherent echo) layer parameters using the
Irkutsk IS radar data. It is shown that our technique has necessary
accuracy (for height and thickness - about 2.5 km, for aspect sensitivity
- 5dB/degree). Processing of the experiments 25-26.12.1998 and 15-16.07.2000
has shown a good agreement of data calculated with the data obtained
by other investigators: an average layer height is 110-120km, average
layer thickness 5km, average aspect sensitivity - 15dB/degree \cite{Foster92}.

The investigation of the experiments with high temporal resolution
allowed us to observe temporal variations of the irregularities layer
parameters. The average thickness and height of the layer does not
contradict the data obtained by other investigators \cite{Haldoupis89,Foster92}.
The investigation of the experiments with high temporal resolution
allowed to detect time variations of the layer parameters. The temporal
variations of the aspect sensitivity are observed by us for the first
time and requires additional investigations.

The work was done under support of RFBR grant \#07-05-01084-a.

\end{document}